\documentclass[fleqn,10pt]{wlscirep}
\usepackage[utf8]{inputenc}
\usepackage[T1]{fontenc}
\usepackage{lineno}
\usepackage{graphicx}
\usepackage{array}
\usepackage{longtable}
\usepackage{multirow}
\usepackage{float}

\title{Scholar Name Disambiguation with Search-enhanced LLM Across Language}

\author[1,$\dag$]{Renyu Zhao}
\author[2,$\dag$]{Yunxin Chen}
\affil[1]{Tencent, SSV, Beijing, 100871, P.R.China}

\affil[*]{corresponding author(s): Renyu Zhao (roizhao@gmail.com)}

\affil[$\dag$]{these authors contributed equally to this work}

\begin{abstract}
    The task of scholar name disambiguation is crucial in various real-world scenarios, including bibliometric-based candidate evaluation for awards, application material anti-fraud measures, and more. Despite significant advancements, current methods face limitations due to the complexity of heterogeneous data, often necessitating extensive human intervention. This paper proposes a novel approach by leveraging search-enhanced language models across multiple languages to improve name disambiguation. By utilizing the powerful query rewriting, intent recognition, and data indexing capabilities of search engines, our method can gather richer information for distinguishing between entities and extracting profiles, resulting in a more comprehensive data dimension. Given the strong cross-language capabilities of large language models(LLMs), optimizing enhanced retrieval methods with this technology offers substantial potential for high-efficiency information retrieval and utilization. Our experiments demonstrate that incorporating local languages significantly enhances disambiguation performance, particularly for scholars from diverse geographic regions. This multi-lingual, search-enhanced methodology offers a promising direction for more efficient and accurate active scholar name disambiguation.

\end{abstract}
\begin{document}

\flushbottom
\maketitle

\thispagestyle{empty}

\section{Introduction}
Entity disambiguation, particularly for scholars' names, is a pivotal task in various contexts such as bibliometric analysis, academic award evaluations, and application material verification. Traditional methods, while advanced, still struggle with the complex, heterogeneous nature of real-world data, leading to performance constraints. Often, these methods require substantial human intervention to achieve satisfactory results. For instance, Ferreira et al.'s "AuthCrowd" \cite{9437769} demonstrated improved author name disambiguation through crowdsourcing, where tasks were broken down for crowd workers, offering a glimpse into the potential of human-assisted disambiguation.

Similarly, Zhang et al. \cite{10.1145/3219819.3219859} implemented a comprehensive framework for name disambiguation within the AMiner system, incorporating both global and local information via a novel representation learning method. They further improved accuracy by integrating human annotators into the disambiguation process. Experimental results on real-world large datasets showed that their proposed solution achieved significantly better performance (7-35\% improvement in F1-score) compared to several state-of-the-art methods. This approach underscores the ongoing necessity and effectiveness of human-in-the-loop methodologies in tackling the disambiguation problem at a large scale.

Despite these advancements, disambiguation tasks remain labor-intensive and prone to inaccuracies, especially in multilingual contexts where the diversity of data sources further complicates the process. The Tsinghua University's KEG lab's "WhoIsWho" dataset exemplifies the challenges faced in resolving authorship based solely on paper metadata, even with the advent of LLMs. The competition aimed to leverage the intelligence of pre-trained models to improve disambiguation accuracy; however, due to the constraints of using only paper metadata, the best solutions \cite{chen2023web} involved fine-tuning LLMs and designing complex comparison processes. These approaches, while effective, did not fully exploit the potential of LLMs.

In practical applications, name disambiguation extends beyond academic papers to include tasks like matching award lists and extracting CV details. These scenarios necessitate robust methods capable of handling varied data forms and linguistic nuances.

To address these challenges, we propose an innovative approach leveraging the robust capabilities of modern search engines combined with LLMs enhanced for multi-language processing. Search engines excel in query rewriting, intent recognition, and data indexing, which can be harnessed to access and compile richer, more nuanced information for disambiguation. This approach acknowledges the importance of local languages in acquiring comprehensive data, particularly for non-native English-speaking scholars. For instance, Chinese scholars may publish extensively in English under various pinyin forms, but their web presence and related information are richer and more informative in Chinese.

Our proposed method integrates these search-enhanced LLMs to construct a more detailed and multi-dimensional profile for each scholar, thereby improving the accuracy and efficiency of name disambiguation tasks. The rest of this paper is structured as follows: Section 2 reviews related work; Section 3 details our methodology; Section 4 presents experimental setups and results; and Section 5 concludes with discussions on implications and future work.

\section{Methodology}

Leveraging the capabilities of LLMs enhanced by search engines forms the core of our methodology. LLMs such as GPT4o and Llama3.2, have demonstrated remarkable proficiency in understanding and generating human language across various contexts \cite{openai2024gpt4technicalreport}. Search-enhanced large language models combine the strengths of pre-trained LLMs with powerful search engine capabilities. While LLMs excel at understanding and generating human-like text across multiple languages, they can be further improved by integrating search functionalities. This enhancement involves utilizing search engines to query external databases or the web, effectively expanding the model's access to real-time, diverse, and dynamic information. By doing so, the model can refine its responses based on the most current and relevant data, leading to more accurate and comprehensive information retrieval. This cross-language capability allows for efficient and effective information utilization, making it especially beneficial for tasks such as scholar name disambiguation, where diverse and heterogeneous data sources are involved.

Integrating LLMs with search engines can significantly enhance their ability to retrieve, interpret, and summarize information from diverse and extensive data sources. This synergy leverages the query rewriting, intent recognition, and data indexing strengths of modern search engines, enabling more accurate and context-aware disambiguation tasks \cite{10.1145/2983323.2983835}. By combining the comprehensive linguistic understanding of LLMs with the vast information retrieval capabilities of search engines, we aim to construct a more detailed and multi-dimensional profile for each scholar, thereby improving the accuracy and efficiency of name disambiguation.

To tackle the problem of scholar name disambiguation, we propose a structured approach that leverages search-enhanced LLMs, which can be regarded as specially designed RAG (Retrival Augmented Generation). The novel approach we propose to scholar name disambiguation leverages the capabilities of search engines and large language models (LLMs). With help of search engine to enhance query rewriting, intent recognition, and data indexing, we can obtain richer information for disambiguation and profile extraction, leading to more comprehensive data dimensions.

\subsection{Module as a Agent}

This methodology we propose is divided into several interconnected modules aimed at improving the precision and efficiency of disambiguating scholars' names. The primary components include an automatic profile extraction pipeline, cross-language name matching, award list matching, and paper name disambiguation. These modules are interdependent, with paper author information extraction and award list name extraction serving as entry points, profiles are compared and paper author names are disambiguated.

To break down, our methodology consists of three main components: scholarly profile extraction, native name retrieval, and profile comparison. Since the core of all the components are LLM, we implemented these components into three LLm agents as follows.

\subsubsection{Extract Agent: Scholarly Profile Extraction}

The Extract Agent is responsible for retrieving relevant literature, profiles, announcements, and other resources using input information. The steps are as follows:

\begin{enumerate}
    \item \textbf{Input Information Parsing}: Parse the input information such as the scholar's name and institution.
    \item \textbf{Search Enhancement}: Utilize search engine capabilities to perform queries with keyword expansion and rewriting to gather more related resources.
    \item \textbf{Information Extraction and Structuring}: Use an LLM to analyze the search results, extract identifiable scholar information (e.g., academic homepages, publication lists, affiliations), and structure this information for further use.
\end{enumerate}

\subsubsection{Name Translate Agent: Native Name Retrieval}

For non-English speaking scholars, particularly Chinese scholars, different forms of pinyin in their names can make it challenging to retrieve complete profiles solely using English information. The Name Translate Agent addresses this by following these steps:

\begin{enumerate}
    \item \textbf{Institution Information Translation}: Partially translate known institution information to enhance search effectiveness.
    \item \textbf{Search Enhancement}: Perform searches based on translated information to retrieve relevant news snippets, academic resumes, etc.
    \item \textbf{Native Name Extraction}: Use an LLM to analyze the search results, identify, and return the native name of the scholar.
\end{enumerate}

\subsubsection{Disambiguation Agent: Profile Comparison}

The Disambiguation Agent compares the information of two scholars with identical names to determine if they refer to the same individual. The steps involved are:

\begin{enumerate}
    \item \textbf{Data Matching}: Initially match different sources of data based on the structured information provided by the Extract Agent.
    \item \textbf{Information Comparison}: Use an LLM to compare detailed contents such as publications, affiliations, co-authors, etc.
    \item \textbf{Decision Making}: Make a final determination on whether the profiles refer to the same scholar based on the comparison results.
\end{enumerate}

\subsection{Workflow Steps}

\begin{figure}[htbp]
    \centering
    \includegraphics[width=0.8\linewidth]{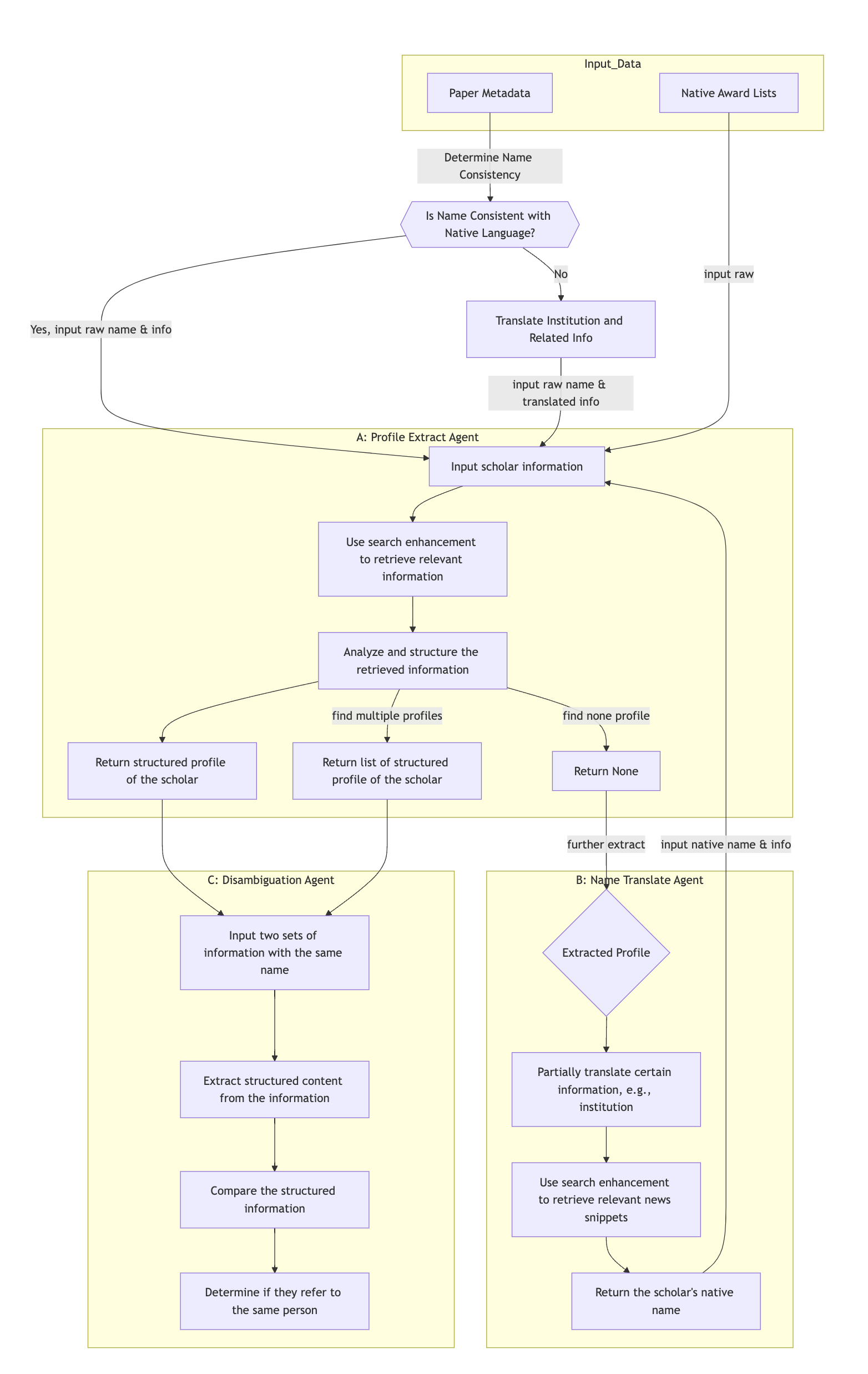}
    \caption{The workflow of our disambiguation method.}
    \label{fig:workflow}
\end{figure}

Figure \ref{fig:workflow} is the visual representation of typical name disambiguation workflows using approach suggest in this paper, logics within each LLM agent are also highlighted in the same block.

\begin{enumerate}
    \item \textbf{Name Consistency Determination}: Assess whether the scholar's name is consistent with their native language.
    \begin{itemize}
        \item \textbf{If Consistent}: Perform a web search using the scholar's name and summarize the homepage and profile using the LLM.
        \item \textbf{If Not Consistent}: Proceed to step 2.
    \end{itemize}
    
    \item \textbf{Institution Translation and Research Area Identification}:
    \begin{itemize}
        \item Use the LLM to translate the institution and related information of the scholar.
        \item Read the paper metadata using the LLM to determine the research area in the author's native language.
        \item Perform a web search with these details combined with the author's name; if possible, summarize the homepage and profile using the LLM.
    \end{itemize}
    
    \item \textbf{Native Name Identification}:
    \begin{itemize}
        \item Use the LLM to infer the scholar's native language name from the search results of step 2.
        \item If the homepage and profile are not found, replace the name in the search keywords with the native language name and search again.
    \end{itemize}
    
    \item \textbf{Multiple Identities Handling}: In cases where search results suggest multiple identities, return a list of potential homepages and profiles for further analysis.
\end{enumerate}

This workflow leverages the query rewriting capabilities of search engines, recognizing that richer scholar-related information often exists on pages in their native languages. By incorporating native language keywords, even when the English name is absent on such pages, relevant information can still be retrieved. If sufficiently rich datasets are available, similar results could be achieved without search engines by indexing all possible name variants.

\subsection{Mathematical Formulation}

Let \( P \) be the set of papers, \( A \) be the set of award lists, and \( S \) be the set of scholars. For paper \( p_i \in P \) with an author list \( \{a_{i1}, a_{i2}, \ldots, a_{im} \} \), and award list \( a_j \in A \) with recipients \( \{r_{j1}, r_{j2}, \ldots, r_{jn} \} \), the objective is mapping \( a_{ik} \) and \( r_{jl} \) to the correct scholar profile \( s \in S \).

Define the following functions:

\[
F_{\text{search}}(q)
\]

\noindent Search function querying the web with query \( q \).

\[
F_{\text{extract}}(\text{result})
\]

\noindent Function to extract profile information from search results.

\[
F_{\text{compare}}(p_1, p_2)
\]

\noindent Function to compare two profiles and assess similarity.

The overall objective is to minimize identity mismatches by optimizing:

\[
\min_{S} \sum_{i,j,k,l} \left| F_{\text{compare}} \left( F_{\text{extract}}(F_{\text{search}}(a_{ik})), F_{\text{extract}}(F_{\text{search}}(r_{jl})) \right) \right|
\]

\noindent where the comparison function aims to identify the correct mappings.

\section{Experiments}

We designed a series of experiments to validate the effectiveness of the proposed method in disambiguating scholar names. The experiments include various search strategies and their impact on recall and precision, compared against baseline methods. Although LLM do relatively easy tasks in these scenarios, yet there might exist some misunderstandings when the given information from webpages are not perfect. Therefore we also made comparisons between two LLMs of gpt4o and Hunyuan in necessary scenarios, so to make additional cross check, and performance evaluations.

\subsection{Baseline: Pinyin + English Institution Name as Input}

Basic web accessable RAG tool could make scholar information extraction given raw information appeared in the papers, romanized names (e.g. Pinyin as for Chinese authors) and their affiliations in English. So we use these to compose prompt for scholar information retrieval, and set the recall rate as baselines. Three search engines (sogou, which is widely used for Chinese search; google, which is world largest one; bing, which is worldwide used while popular in Chinese search) and two LLMs (OpenAI gpt4o and Tencent hunyuan) are used for more comprehensive comparison.

Although search engine has many merits for name disambiguation as metioned before, since web pages retrieved from search engines often contain substantial news articles about scholars, however, simple scholar introductions are not always useful information. To effectively extract relevant scholar information, we designed the following steps:

\begin{enumerate}
    \item The model first needs to determine if the web content contains detailed biographical information about the scholar, such as educational background, work experience, and research areas.
    \item Next, the model must assess whether the web information is related to the target scholar, as search results may include other scholars with the same name or from the same institution. This issue may arise due to common names or the search engine prioritizing partial keyword matches when it cannot fully cover the prompt. For example, searching for "Lin Jing, Sun Yat-sen University" might return information related to "Lin Liang" instead because of incorrect data collection or misrecognition of Chinese names.
    \item Finally, the remaining web information is deduplicated by pairwise comparison to eliminate redundant information. This step is crucial because large institutions might have multiple scholars with the same name, and insufficiently detailed institution names could result in multiple matching scholars.
\end{enumerate}

Table \ref{table:baseline_result} show the baseline reuslts.

\begin{table}[htbp]
    \centering
    \begin{tabular}{| p{5cm} | p{2cm} | p{2cm} | p{2cm} |}
      \hline
      \textbf{Pinyin+Institution\_en} & \textbf{sogou} & \textbf{google} & \textbf{bing} \\
      \hline
        gpt4o & 40\% & 44\% & 49\% \\
        hunyuan & 15\% & 22\% & 31\% \\
      \hline
    \end{tabular}
    \caption{Basic RAG Recall Result}
    \label{table:baseline_result}
\end{table}

\subsection{Pinyin + Chinese Institution Name}

As we all aware that search engine prone to return webpages of matching language as search query, a simple strategy to improve basic RAG is to translate scholars' affiliations into its native language name, which is trival yet effective. 

Experiment result in Table \ref{table:simple_institution_trans_result} shows that, using a combination of Pinyin and the Chinese institution in a Chinese search engine yields useful information for a few scholars whose homepages contain their Pinyin name, either in their profiles or in their publication lists. However, there are still many cases when search engines fail to accurately capture Pinyin keywords, making the search results more focused on the institution's information rather than the scholar's details. 

\begin{table}[htbp]
    \centering
    \begin{tabular}{| p{5cm} | p{2cm} | p{2cm} | p{2cm} |}
      \hline
      \textbf{Pinyin+Institution\_zh} & \textbf{sogou} & \textbf{google} & \textbf{bing} \\
      \hline
        gpt4o & 60\% & 57\% & 61\% \\
        hunyuan & 30\% & 41\% & 38\% \\
      \hline
    \end{tabular}
    \caption{RAG Recall Result with Institution Translated}
    \label{table:simple_institution_trans_result}
\end{table}

Table \ref{table:researchers_info_sample} show sample output format, and the first line is result of this approach.

\begin{table}[htbp]
    \centering
    \begin{tabular}{| p{2cm} | p{5cm} | p{9cm} |}
      \hline
      \textbf{Name} & \textbf{Affiliation} & \textbf{Extract Agent Output} \\ [0.5ex] 
      \hline
        Zhang, Yihui & Tsinghua University, Center for Flexible Electronics Technology, Department of Engineering Mechanics, Applied Mechanics Lab, Beijing 100084, People's Republic of China & 
        \{"name": "Zhang Yihui", "workplace": "Tsinghua University Center for Flexible Electronics Technology, Jiaxing Zhejiang Tsinghua Flexible Electronics Technology Research Institute", "keywords": \verb|[|"Flexible electronic devices and manufacturing", "Flexible microsystem and heterogeneous integration technology", "Integration technology of flexible electronics and platform", "Wearable technology and human-machine hybrid intelligence"\verb|]|, "education\_track": \verb|[|\verb|]| , "professional\_track": \verb|[|\verb|]| , "honor\_track": \verb|[|\verb|]| \} \\
      \hline
        Zhang, Yihui & Tsinghua University, Center for Flexible Electronics Technology, Department of Engineering Mechanics, Applied Mechanics Lab, Beijing 100084, People's Republic of China &
        \{"name": "Zhang Yihui", "workplace": "Center for Flexible Electronics Technology, Tsinghua University", "keywords": \verb|[| "3D Micro/Nano Structure Assembly", "Soft Matter and Flexible Structural Mechanics", "Microrobots"\verb|]|, "education\_track": \verb|[|\verb|]|, "professional\_track": \verb|[| \{"fromto": "null", "agency": "Tsinghua University", "title": "Tenured Professor"\}\verb|]|, "honor\_track": \verb|[| \{"time": "null", "award": "National Science Fund for Distinguished Young Scholars"\}, \{"time": "null", "award": "Science Exploration Award"\}, \{"time": "null", "award": "Gustus L. Larson Memorial Award"\}, \{"time": "null", "award": "MIT TR35"\}, \{"time": "null", "award": "ASME Thomas J.R. Hughes Young Investigator Award"\}\verb|]|\} \\
    \hline
    \end{tabular}
    \caption{Researchers' Information Example (translated back to English, jsonified)}
    \label{table:researchers_info_sample}
\end{table}

\subsection{Chinese Name + Institution Name}

To further retrieve scholars' information for comparison, as we analyzed before, it is worthy to retrieve scholars name in native language with a sophisticatedly designed agent, then apply to the Information Retrieval Agent again.

Using Pinyin and the primary institution in search engine to get scholar homepages or news reports allows us to infer the corresponding Chinese name. Combining this inferred Chinese name with the Chinese institution for further search enhanced retrieval will significantly improves the recall.

\subsubsection{Retrieving Chinese Names for Chinese Authors}

Despite occasional inaccuracies in the inferred Chinese names, as long as the transliteration is correct, the model effectively retrieves target information. This strategy shows a correct Chinese name identification rate of 87\% and a recall rate of 74\%.

Tables summarizes the recall and precision rates for each strategy to get Chinese scholars' native name are listed in this paper.

\begin{table}[htbp]
    \centering
    \begin{tabular}{| p{5cm} | p{2cm} | p{2cm} | p{2cm} |}
      \hline
      \textbf{Pinyin+Institution\_zh} & \textbf{sogou} & \textbf{google} & \textbf{bing} \\
      \hline
        gpt4o & 48\% & 72\% & 87\% \\
        hunyuan & 41\% & 67\% & 80\% \\
      \hline
    \end{tabular}
    \caption{Recall of Chinese Scholars' Native Name}
    \label{table:chinese_name_recall}
\end{table}

\begin{table}[htbp]
    \centering
    \begin{tabular}{| p{5cm} | p{2cm} | p{2cm} | p{2cm} |}
      \hline
      \textbf{Pinyin+Institution\_zh} & \textbf{sogou} & \textbf{google} & \textbf{bing} \\
      \hline
        gpt4o & 44\% & 52\% & 74\% \\
        hunyuan & 38\% & 54\% & 69\% \\
      \hline
    \end{tabular}
    \caption{Precision of Chinese Scholars' Native Name}
    \label{table:chinese_name_precision}
\end{table}

And second row of Table \ref{table:researchers_info_sample} show that with this strategy, scholars with few information in the simple "Pinyin + Chinese Institution Name" get more detailed and thorough.

E-mails of authors appeared in papers, could somehow as a supplement for retrieve their native names, even homepages. Using Pinyin, Chinese institution, and contact information in Chinese search engines lets the model infer the Chinese name, then combine it with the Chinese institution for further searches. Although the proper recognition rate is slightly lower than from Google results, this strategy serves as a supplement when previous methods fall short.

As shown in \ref{table:chinese_name_precision_email} and \ref{table:chinese_name_recall_email}, in this approach, the best Chinese name identification rate is boosted to 80\% and a recall rate to 95\%.

\begin{table}[htbp]
    \centering
    \begin{tabular}{| p{5cm} | p{2cm} | p{2cm} | p{2cm} |}
      \hline
      \textbf{Pinyin+Institution\_zh} & \textbf{sogou} & \textbf{google} & \textbf{bing} \\
      \hline
        gpt4o & 47\% & 95\% & 93\% \\
        hunyuan & 42\% & 84\% & 92\% \\
      \hline
    \end{tabular}
    \caption{Recall of Chinese Scholars' Native Name with Email}
    \label{table:chinese_name_recall_email}
\end{table}

\begin{table}[H]
    \centering
    \begin{tabular}{| p{5cm} | p{2cm} | p{2cm} | p{2cm} |}
      \hline
      \textbf{Pinyin+Institution\_zh} & \textbf{sogou} & \textbf{google} & \textbf{bing} \\
      \hline
      gpt4o & 23\% & 61\% & 80\% \\
      hunyuan & 27\% & 66\% & 78\% \\
      \hline
    \end{tabular}
    \caption{Precision of Chinese Scholars' Native Name with Email}
    \label{table:chinese_name_precision_email}
\end{table}

Table \ref{table:researchers_info_sample_email} show that with email added in native name retrieval, scholar's information could be fetched by our agent in difficult tasks.

    \begin{table}[h]
        \centering
        \begin{tabular}{| p{2cm} | p{4cm} | p{10cm} |}
            \hline
            \textbf{With Email} & \textbf{Name and Affiliation} & \textbf{Extract Agent Output} \\ [0.5ex] 
    \hline
      No & \multirow{2}{4cm}{Qiang, Zhimin, University of Chinese Academy of Sciences, Chinese Academy of Sciences, Research Center for Eco-environmental Sciences, Key Laboratory of Drinking Water Science \& Technology, 18 Shuangqing Rd, Beijing 100085, People's Republic of China} & "null" \\
      \cline{1-1}
      \cline{3-3}
      Yes &  & 
      \{"name": "Qiang Zhimin", "workplace": "Research Center for Eco-environmental Sciences, Chinese Academy of Sciences", "email": \verb|[|"qiangz@rcees.ac.cn"\verb|]|, "keywords": \verb|[|"Trace Micro-pollutants in Water", "Wastewater Treatment", "Drinking Water Purification", "Migration and Control of Emerging Contaminants in Water Treatment Processes", "Polluted River Restoration"\verb|]|, "education\_track": \verb|[|\{"fromto": "1994 - 1997", "school": "Tongji University, Department of Environmental Engineering", "major": "null", "scholar": "Master"\}, \{"fromto": "2002", "school": "University of Delaware, USA", "major": "null", "scholar": "PhD"\}\verb|]|, "professional\_track": \verb|[|\{"fromto": "2001 - 2006", "agency": "Missouri University of Science and Technology, USA", "title": "Postdoctoral Fellow"\}, \{"fromto": "2001 - 2006", "agency": "Missouri University of Science and Technology, USA", "title": "Assistant Research Professor"\}, \{"fromto": "2001 - 2006", "agency": "Missouri University of Science and Technology, USA", "title": "Associate Research Professor"\}, \{"fromto": "2006 - Present", "agency": "Research Center for Eco-environmental Sciences, Chinese Academy of Sciences", "title": "null"\}\verb|]|, "honor\_track": \verb|[|\verb|]|\} \\
    \hline
\end{tabular}
\caption{Study of Email Used in Search Enhancement Results (translated back to English, jsonified)}
\label{table:researchers_info_sample_email}
\end{table}

Overall, with Chinese authors' native name retrieved, the best result we achieve to get the scholars' information reaches 96\%, full comparisons list in Table \ref{table:chinese_name_result}

\begin{table}[H]
    \centering
    \begin{tabular}{| p{5cm} | p{2cm} | p{2cm} | p{2cm} |}
      \hline
      \textbf{Pinyin+Institution\_zh} & \textbf{sogou} & \textbf{google} & \textbf{bing} \\
      \hline
      gpt4o & 93\% & 89\% & 96\% \\
      hunyuan & 90\% & 92\% & 94\% \\
      \hline
    \end{tabular}
    \caption{Overall Recall of Scholars Information with their Native Name if possible}
    \label{table:chinese_name_result}
\end{table}

\subsection{Scholar Name Disambiguation with Retrieved Information}

To compare scholars, we assign scores to various key fields: matching institutions earn 2 points, each repeated educational or work experience segment earns 3 points, and research area keywords earn 1-4 points depending on relevance. The decision threshold is calculated as the sum of the institution score, one educational or work segment score, and the median keyword score. Thus, the boundary value is set at 7 points. Scholars whose similarity scores meet or exceed 7 points are considered the same individual, otherwise they are deemed different. Human experts evaluation assisted with cross validation yields that the result in these dataset is can almost achieve accuracy of human expert as shown in Table \ref{table:name_disambiguation_result}.

\begin{table}[H]
    \centering
    \begin{tabular}{| p{5cm} | p{5cm}|}
    \hline
        Model & Accuracy \\
    \hline
        hunyuan & 86.6\% \\
        gpt4o & 98\% \\
    \hline
    \end{tabular}
    \caption{Accuracy of Name Disambiguation with Retrieved Scholar Information}
    \label{table:name_disambiguation_result}
\end{table}

Through these experiments, we validate our proposed search-enhanced LLM-based approach for scholar name disambiguation. The results demonstrate that our method significantly improves disambiguation precision and efficiency across different language environments, particularly excelling in disambiguating Chinese scholars.


\section*{Usage Notes}

The dataset provided in this study includes publicly available information sourced from web pages, specifically focusing on author details from other academic publications. This data is intended solely for academic and research purposes to support reproducibility and transparency of scientific research.

Researchers using this dataset are encouraged to respect the original context and purpose of the data. Any secondary use of the dataset should adhere to ethical guidelines and best practices in data handling, including proper citation of the original sources.

While the data included in this dataset is publicly accessible, it is important to acknowledge potential privacy considerations. The collection, use, and dissemination of this information have been conducted with the intent to respect privacy and intellectual property rights.

Users of this dataset must ensure compliance with relevant data protection regulations, institutional policies, and ethical standards. Neither the authors of this study nor their affiliated institutions assume responsibility for any misuse of the data or any legal implications that may arise from its use. It is the responsibility of the end-users to ensure that their usage of the dataset does not infringe upon the privacy or rights of individuals.

\section*{Code availability}
Data and code, including prompts used in build the agents are available at following github project: 
\href{https://github.com/sylviachen0513/Scholar\_Name\_Disambiguation\_with\_Search\_enhanced\_LLM\_Across\_Language}{https://github.com/} \\

\bibliography{ref}

\section*{Author contributions statement}
R.Z. conceived the experiment(s), Y.C. and R.Z. conducted the experiment(s). All authors reviewed the manuscript. 

\section*{Competing interests}

The corresponding author is responsible for providing a \href{https://www.nature.com/sdata/policies/editorial-and-publishing-policies#competing}{competing interests statement} on behalf of all authors of the paper.



\end{document}